# Thermal Energy Budget of Electrons in the Inner Heliosphere: Parker Solar Probe Observations

Joel B. Abraham[1], Daniel Verscharen[1], Robert T. Wicks[2], Jeffersson A. Agudelo Rueda[3], Christopher J. Owen[1], Georgios Nicolaou[1], and Seong-Yeop Jeong[2]
[1] Mullard Space Science Laboratory, University College London, Dorking, Surrey, RH5 6NT, UK; joel.abraham.19@ucl.ac.uk
[2] Department of Mathematics, Physics and Electrical Engineering, Northumbria University, Newcastle upon Tyne, NE1 8ST, UK
[3] Department of Physics and Astronomy, Dartmouth College, Hanover, NH 03755, USA
Received 2022 September 27; revised 2022 November 1; accepted 2022 November 1; published 2022 December 19

## Abstract

We present an observational analysis of the electron thermal energy budget using data from Parker Solar Probe. We use the macroscopic moments, obtained from our fits to the measured electron distribution function, to evaluate the thermal energy budget based on the second moment of the Boltzmann equation. We separate contributions to the overall budget from reversible and irreversible processes. We find that an irreversible thermal energy source must be present in the inner heliosphere over the heliocentric distance range from 0.15 to 0.47 au. The divergence of the heat flux is positive at heliocentric distances below 0.33 au, while beyond 0.33 au, there is a measurable degradation of the heat flux. Expansion effects dominate the thermal energy budget below 0.3 au. Under our steady-state assumption, the free streaming of the electrons is not sufficient to explain the observed thermal energy density budget. We conjecture that the most likely driver for the required heating process is turbulence. Our results are consistent with the known nonadiabatic polytropic index of the electrons, which we measure as 1.18 in the explored range of heliocentric distances.

*Unified Astronomy Thesaurus concepts:* The Sun (1693); Solar wind (1534); Heliosphere (711); Plasma physics (2089)

## 1. Introduction

The solar wind is a highly ionized plasma that continuously flows out from the solar corona and fills the heliosphere with protons, electrons, $\alpha$-particles, and small traces of heavier ions (Marsch 2006). Electrons evolve in a complex manner as they propagate into the heliosphere under the influence of processes such as expansion, turbulent dissipation, plasma instabilities, wave–particle interactions, Coulomb collisions, global electric fields, and gravity (Lie-Svendsen et al. 1997; Vocks & Mann 2003; Smith et al. 2012; Jeong et al. 2022a). Each of these processes may play a role in modifying the shape of the electron velocity distribution function (VDF), which in turn has significant ramifications for the solar wind energy budget.

Electrons have a strong impact on the solar wind thermodynamics. Due to their small mass compared to the ions, electrons make negligible contributions to the solar wind's total mass, momentum, and kinetic-energy fluxes. Due to their larger thermal speeds, however, electrons play a key role in influencing the thermal energy budget by efficiently carrying away heat from the Sun (Cranmer et al. 2009; Landi et al. 2014; Štverák et al. 2015; Halekas et al. 2021). Most kinetic models of solar wind acceleration rely on a contribution from the electric field, which arises due to ambipolar diffusion as a consequence of the strong electron pressure gradient between the corona and the heliosphere (Lemaire & Scherer 1971; Parker 2010).

The electron VDF is often reported to be composed of three populations: the core, the halo, and the strahl (Feldman et al. 1975; Maksimovic et al. 2005; Štverák et al. 2009; Abraham et al. 2022). The thermal core population usually takes a Maxwellian form and comprises more than 90% of the total local electron density. The halo population is present at higher energies and is often modeled as a bi-$\kappa$-distribution. The core and halo are each quasi-isotropic and thus have significant electron fluxes at all pitch angles. The strahl takes the form of a magnetic-field-aligned beam of electrons moving parallel or antiparallel (or sometimes bidirectionally) to the local magnetic field.

In situ observations of the solar wind electron temperature gradients often show deviations from adiabatic profiles but vary between adiabatic and isothermal expectations (Ogilvie & Scudder 1978; Feldman et al. 1979; Pilipp et al. 1990; Maksimovic et al. 2000). This observation motivates the important need to explain the physical mechanisms responsible for the nonadiabatic evolution of the plasma electrons. To date, few studies have addressed the nonadiabatic behavior of the electron temperature. However, Pilipp et al. (1990), Štverák et al. (2008), and Cranmer et al. (2009) have quantified the heating/cooling rates required in the solar wind to support the observed radial profiles of the temperature and the heat flux.

Pilipp et al. (1990) provide a basic formulation of the electron energetics. An observed fast stream reported in this study is shown to require external heating, while an observed slow stream does not need external heating at distances between 0.3 and 1 au. However, due to the limited data set used in this study, assessment of the full global evolution of the thermal energy budget was not possible. More recently, work on the energy budget is presented by Štverák et al. (2015). Through an approximation of the thermal energy balance, this study finds that no external heat mechanisms are required to explain the electron temperature profile observed in slow solar wind between 0.3 and 1 au. The required local energy input is provided by degradation of the electron heat flux. In addition, this study highlights that internal electron energy is transported from the parallel to the perpendicular degrees of freedom.

In this paper, we evaluate the electron thermal energy balance by first providing and examining the exact and complete description of the thermal energy density based on the second moment of the Boltzmann equation. We address the fundamental question about the deposition rate of thermal energy as a function of heliocentric distance in the very inner heliosphere. We quantify the contributions of thermal energy sinks and sources using the Parker Solar Probe (PSP) data set.

In Section 2, we derive a framework based on the second moment of the Boltzmann equation to study the conservation of thermal energy. This approach is based on the theoretical framework developed by Agudelo Rueda et al. (2022) and Agudelo Rueda (2022). In Section 3, we apply this framework to our PSP data set described by Abraham et al. (2022). In Section 4, we present our results. In Section 5, we discuss our results and the implications of this work. Finally, in Section 6, we summarize our findings and outline perspectives for future work.

## 2. Thermal Energy Budget

The evolution of the VDF in phase space follows the Boltzmann equation

$$\frac{\partial f}{\partial t} + \boldsymbol{v} \cdot \nabla f + \frac{q}{m}(\boldsymbol{E} + \boldsymbol{v} \times \boldsymbol{B}) \cdot \nabla_v f = \left(\frac{df}{dt}\right)_{\text{Coll}}, \quad (1)$$

where $f$ is the VDF, $t$ is the time, $\boldsymbol{v}$ is the velocity, $q$ is the charge of a particle, $m$ is the mass of a particle, $\boldsymbol{E}$ is the electric field, and $\boldsymbol{B}$ is the magnetic field. The term $(df/dt)_{\text{Coll}}$ represents the change in the distribution function due to collisions.

To address the thermal energy budget of the electrons, we take the second moment of Equation (1) for electrons, which leads to the following differential equation:

$$\frac{\partial \epsilon}{\partial t} + (\boldsymbol{u}_e \cdot \nabla)\epsilon + \nabla \cdot \boldsymbol{Q}_e + \nabla \boldsymbol{u}_e : \mathsf{P}_e + (\nabla \cdot \boldsymbol{u}_e)\epsilon = \Xi, \quad (2)$$

where $\epsilon$ is the thermal energy density, $\boldsymbol{Q}_e$ is the electron heat flux, $\boldsymbol{u}_e$ is the electron bulk velocity, $\mathsf{P}_e$ is the electron pressure tensor, and $\Xi$ is the irreversible contribution to the thermal energy budget (i.e., the second moment of $(df/dt)_{\text{Coll}}$). The thermal energy density $\epsilon$ is defined as

$$\epsilon = \frac{1}{2}\text{Tr}(\mathsf{P}_e), \quad (3)$$

where $\text{Tr}(\mathsf{P}_e)$ is the trace of the electron pressure tensor.

The left-hand side of Equation (2) is the reversible transfer of the thermal energy density, and the right-hand side is the irreversible transfer of the thermal energy density. On the left-hand side, the first term describes the partial time derivative of the thermal energy density. The second term describes the advection of $\epsilon$ with $\boldsymbol{u}_e$. The third term describes the divergence of the electron heat flux. The fourth term describes the pressure strain interaction, and the fifth term describes the impact of expansion/contraction of the solar wind plasma. The right-hand side represents any spatial deposition of thermal energy through irreversible processes. If the right-hand side is zero, then the energy budget of the measured electron distribution evolves reversibly. If the right-hand side is nonzero, then the thermal energy budget of the electron distribution has a sink or a source of thermal energy density. The quantity $\Xi$ thus corresponds to the irreversible thermal power density of the electrons.

We transform Equation (2) into a usable format that can take the fit parameters of our electron VDF fits as inputs. As we use data from a single spacecraft, it is not possible to measure the temporal evolution of the plasma for a given point in space, as this would require simultaneous multispacecraft measurements. To account for this, we assume steady-state conditions, which removes the partial time derivative in Equation (2). Therefore, all remaining terms on the left-hand side of Equation (2) arise from free-streaming effects in Equation (1).

### 2.1. Advection of the Thermal Energy Density by Bulk Flows

Assuming spherical symmetry, we write the second term of Equation (2) as

$$(\boldsymbol{u}_e \cdot \nabla)\epsilon = u_r \frac{\partial \epsilon}{\partial r}, \quad (4)$$

where $u_r$ represents the radial electron bulk speed.

In our data analysis, we use the radial proton speed instead of the radial electron speed for $u_r$ because of the technical difficulty in determining the electron radial speed from data directly. The proton radial speed is approximately the same as the electron bulk speed due to the zero-current requirement.

We derive $\epsilon$ from the second moment of the VDF, which defines the pressure tensor:

$$\mathsf{P}_e = m_e \int f_e (\boldsymbol{v} - \boldsymbol{u}_e)(\boldsymbol{v} - \boldsymbol{u}_e) d^3v, \quad (5)$$

where $m_e$ is the mass of the electron and $f_e$ is the electron distribution function. Our fit results provide densities, bulk velocities, and pressure tensors for the three electron populations separately (Abraham et al. 2022). We now derive total electron moments from these quantities. Integrating and solving for Equation (5) in terms of each electron population, we obtain the following:

$$\mathsf{P}_e = \mathsf{P}_c + \mathsf{P}_h + \mathsf{P}_s + m_e[n_s(\boldsymbol{u}_s - \boldsymbol{u}_e)^2 \\ + n_c(\boldsymbol{u}_c - \boldsymbol{u}_e)^2 + n_h(\boldsymbol{u}_h - \boldsymbol{u}_e)^2]\hat{\boldsymbol{b}}\hat{\boldsymbol{b}}, \quad (6)$$

where $n_c$ is the core density, $n_h$ is the halo density, $n_s$ is the strahl density, $\boldsymbol{u}_c$ is the core bulk speed, $\boldsymbol{u}_h$ is the halo bulk speed, $\boldsymbol{u}_s$ is the strahl bulk speed, and $\hat{\boldsymbol{b}}$ is the unit vector of the magnetic field.

The total pressure arises from the thermal pressures of each population and from the drifts of each population with respect to the electron bulk speed. Taking the trace of the pressure tensor from Equation (6), we obtain

$$\text{Tr}(\mathsf{P}_e) = \sum_{j=c,h,s} (n_j k_b (2T_{\perp j} + T_{\| j}) + n_j m_e (u_j - u_e)^2), \quad (7)$$

where the subscript $j$ indicates the core ($c$), halo ($h$), and strahl ($s$) populations; $k_b$ is the Boltzmann constant; $T_{\perp j}$ is the perpendicular temperature; $T_{\| j}$ is the parallel temperature; and $u_j$ is the bulk speed of population $j$.

We decompose the total electron bulk speed as

$$\boldsymbol{u}_e = \frac{1}{n_e} \sum_j n_j \boldsymbol{u}_j, \quad (8)$$

where $n_e = n_c + n_h + n_s$ is the total electron density. Substituting Equation (7) into Equation (4), we obtain

$$(\boldsymbol{u}_e \cdot \nabla)\epsilon = \frac{1}{2} u_r \frac{\partial}{\partial r} \sum_{j=c,h,s} [n_j k_b (2T_{\perp j} + T_{\|j}) + n_j m_e (\boldsymbol{u}_j - \boldsymbol{u}_e)^2] \quad (9)$$

### 2.2. Divergence of Heat Flux

The third term in Equation (2) describes the divergence of the electron heat flux. The divergence of the heat flux can be expressed as

$$\nabla \cdot \boldsymbol{Q}_e = \frac{1}{r^2} \frac{\partial (r^2 Q_r)}{\partial r}, \quad (10)$$

where $Q_r$ is the radial electron heat flux. The heat flux vector is defined as the third moment of the VDF:

$$\boldsymbol{Q}_e = \frac{m_e}{2} \int f_e (\boldsymbol{v} - \boldsymbol{u}_e) \cdot (\boldsymbol{v} - \boldsymbol{u}_e)(\boldsymbol{v} - \boldsymbol{u}_e) d^3 v. \quad (11)$$

Integrating Equation (11) yields

$$Q_r = \sum_j [Q_{rj} + \tfrac{3}{2} n_j k_b T_{\|j} (\boldsymbol{u}_j - \boldsymbol{u}_e) + (\boldsymbol{u}_j - \boldsymbol{u}_e) n_j k_b T_{\perp j}$$
$$+ \tfrac{1}{2} m_e n_j (\boldsymbol{u}_j - \boldsymbol{u}_e)^3] \hat{\boldsymbol{b}}_r, \quad (12)$$

where $Q_{rj}$ represents the intrinsic heat flux of population $j$ that arises due to reflectional asymmetries in the individual population's distribution around its bulk speed. For symmetric component VDFs, this term vanishes. Substituting Equation (12) into Equation (10), we obtain the following expression for the divergence of the heat flux in terms of the fit parameters:

$$\nabla \cdot \boldsymbol{Q}_e = \tfrac{1}{r^2} \tfrac{\partial}{\partial r} r^2 \sum_j [\tfrac{3}{2} n_j k_b T_{\|j} (\boldsymbol{u}_j - \boldsymbol{u}_e)$$
$$+ (\boldsymbol{u}_j - \boldsymbol{u}_e) n_j k_b T_{\perp j} + \tfrac{1}{2} m_e n_j (\boldsymbol{u}_j - \boldsymbol{u}_e)^3] \hat{\boldsymbol{b}}_r. \quad (13)$$

### 2.3. Pressure Strain Term

The fourth term in Equation (2) is called the pressure strain term. This term quantifies a contribution to the thermal energy density by pressure that is advected by spatially inhomogeneous bulk flows. Yang et al. (2022) show that the pressure strain term serves as an estimate of the transfer of bulk-flow energy into thermal energy. Therefore, the pressure strain term is an important tracer for changes in internal energy.

Using Equation (6), we obtain the following expression for the pressure strain term:

$$\nabla \boldsymbol{u}_e : \mathsf{P}_e = \sum_j (\nabla \boldsymbol{u}_j : \mathsf{P}_j + \nabla \boldsymbol{u}_j : m_e [n_j (\boldsymbol{u}_j - \boldsymbol{u}_e)^2] \hat{\boldsymbol{b}} \hat{\boldsymbol{b}}). \quad (14)$$

We fit our data in the magnetic-field-aligned coordinate system ($V_\|$, $V_{\perp_1}$, $V_{\perp_2}$). For our analysis of Equation (2), we work in the spherical coordinate system (radial distance $r$, polar angle $\theta$, and azimuth angle $\phi$). To account for this, we transform the pressure tensor from the magnetic-field-aligned frame to the spherical coordinate frame by applying the following transformation:

$$T^{-1} \tilde{\mathsf{P}}_e T = \mathsf{P}_e, \quad (15)$$

where $T$ is the transformation matrix, $T^{-1}$ is the inverse of the transformation matrix, $\tilde{\mathsf{P}}_e$ is the pressure tensor in the magnetic-field-aligned frame, and $\mathsf{P}_e$ is the pressure tensor in the spherical coordinate frame. The transformation matrix is defined as follows with the assumptions $u_\phi \neq 0$, $u_\theta = 0$, $\partial/\partial\theta = 0$, and $\partial/\partial\phi = 0$:

$$T = \begin{pmatrix} a_1 b_T & -a_2 b_T & -a_3 \\ a_2 & a_1 & 0 \\ b_R & -b_N & b_T \end{pmatrix}, \quad (16)$$

where

$$a_1 = \frac{B_R}{\sqrt{B_R^2 + B_N^2}}, \quad a_2 = \frac{B_N}{\sqrt{B_R^2 + B_N^2}}, \quad a_3 = \frac{\sqrt{B_R^2 + B_N^2}}{|B|},$$
$$b_R = \frac{B_R}{|B|}, \quad b_T = \frac{B_T}{|B|}, \quad b_N = \frac{B_N}{|B|}. \quad (17)$$

$B_R$ is the magnetic field component in the radial direction, $B_T$ is the magnetic field component in the tangential direction, and $B_N$ is the magnetic field component in the normal direction. Applying the transformation in Equation (15), we rewrite Equation (14) as

$$\nabla \boldsymbol{u}_e : \mathsf{P}_e = \frac{\partial u_r}{\partial r} \mathsf{P}_{e,rr} + \frac{u_r}{r} (\mathsf{P}_{e,\theta\theta} + \mathsf{P}_{e,\phi\phi}) + \frac{\partial u_\phi}{\partial r} \mathsf{P}_{e,r\phi}$$
$$- \frac{u_\phi}{r} \left( \mathsf{P}_{e,r\phi} + \frac{\mathsf{P}_{e,\theta\phi}}{\tan\theta_B} \right), \quad (18)$$

where $\theta_B$ is the angle between the magnetic field direction and the radial direction.

### 2.4. Expansion Rate

The fifth term in Equation (2) quantifies the effects of expansion and compression on the thermal energy density:

$$(\nabla \cdot \boldsymbol{u}_e)\epsilon = \frac{1}{r^2} \frac{\partial (r^2 u_r)}{\partial r} \epsilon. \quad (19)$$

This term arises directly from the divergence of the electron bulk velocity in the expanding solar wind.

### 2.5. Polytropic Index

Although our kinetic description in Equation (1) does not require a polytropic closure, it is useful to compare our results with the 1olytropic properties of the electron fluid in the solar wind. The polytropic relationship describes the correlation between the pressure and the density of a fluid during a transition from one state to another. During a polytropic process, the ratio between the energy transferred as heat to the energy done as work is constant (Parker 1963; Chandrasekhar 1967). For an electron plasma, the polytropic relationship is given as

$$P_e \propto n_e^{\gamma_e}, \quad (20)$$

where $P_e$ is the scalar electron pressure and $\gamma_e$ is the electron polytropic index. The polytropic relationship brings closure to the moments hierarchy (e.g., Kuhn et al. 2010) as it relates the second-order moment (pressure or temperature) with the zeroth-order moment (density). Moreover, the value of the polytropic index is a useful metric to investigate the nature of the mechanisms in the solar wind electrons that involve heat transfer.

By taking the logarithm of Equation (20), we obtain

$$\log(P_e) = \gamma_e \log(n_e) + C, \quad (21)$$

where $C$ is a constant. Equation (21) allows us to determine $\gamma_e$ from linear fits to observations of $\log(P_e)$ versus $\log(n_e)$. The slope of the fitted polytropic model determines $\gamma_e$ (Totten et al. 1995; Nicolaou et al. 2020). An adiabatic electron fluid exhibits $\gamma_e = 5/3$, and an isothermal electron fluid exhibits $\gamma_e = 1$. We calculate the scalar pressure as $P_e = \text{Tr}(\mathsf{P}_e)/3$.

## 3. Parker Solar Probe Data

PSP was launched in 2018 August and will achieve the closest perihelion at a heliocentric distance of 9.86 solar radii in 2024. We analyze the electron VDFs measured by the Solar Wind Electrons, Alphas, and Protons (SWEAP) instrument suite on board PSP (Kasper et al. 2016). The SWEAP instrument measures the 3D electron VDF with the Solar Probe Analyzer-Electron (SPAN-E) sensor, consisting of two top-hat electrostatic analysers: SPAN-A and SPAN-B. Together, the two electrostatic analyzers measure electrons arriving from across almost the full sky using orthogonally positioned $120° \times 240°$ fields of view, over an energy range from 2 to 1793 eV. A detailed description of the SWEAP instrument and the operational modes is given by Whittlesey et al. (2020).

We use the level-3 pitch-angle data product. The level-3 data are provided in 32 energy bins and in 12 pitch-angle bins of width 15° with bin centers ranging from 7°.5 to 172°.5. The proton bulk velocity is obtained from SWEAP's Solar Probe Cup (SPC) sensor (Case et al. 2020).

We use the fitted parameters and the solar wind speeds obtained by Abraham et al. (2022). We use the magnetic field data provided in the level-3 data product. The core is fitted with a bi-Maxwellian distribution, the halo with a bi-$\kappa$-distribution, and the strahl with a drifting bi-Maxwellian distribution. We fit the sum of the three populations to capture the total envelope of the measured distribution. We evaluate the goodness of the fit using the reduced $\chi$-squared parameter as described by Abraham et al. (2022).

### 3.1. Data Handling

We use the fitted parameters directly to calculate all the terms in Equations (9), (13), (18), and (19). We then split the data into 10 radial bins and use the central difference method to approximate the necessary radial gradients. We account for the statistical uncertainties in each term in Equations (9), (13), (18), and (19) by propagating the errors of the fitted parameters using the Monte Carlo error propagation technique, which we evaluate for 100,000 iterations. For each radial distance bin, we plot the mean and the confidence interval as the standard deviation of the sorted samples from the Monte Carlo method.

## 4. Results

In Figure 1, we show the four contributions to the left-hand side of Equation (2). The blue line represents the advection due to the bulk flow, the red line represents the divergence of the heat flux, the yellow line represents the pressure strain term, and the purple line represents the expansion term. The error bars for each term decrease with radial distance.

On a linear scale, it is difficult to recognize the relative contributions of each term at larger radial distances. Therefore, we plot the means of these terms on a logarithmic scale in Figure 2. When a contribution is positive, we use a solid line. When a contribution is negative, we use a dashed line. We also plot the resulting irreversible electron thermal power density $\Xi$ from the right-hand side of Equation (2). We propagate the cumulative errors to $\Xi$. The $\Xi$ term decreases from $\sim 1 \times 10^{-13}$ W m$^{-3}$ at 0.15 au to $\sim 3 \times 10^{-15}$ W m$^{-3}$ at 0.35 au before being roughly constant at $\sim 3 \times 10^{-15}$ W m$^{-3}$ within the error bars from 0.35 to 0.48 au. The advection, pressure strain, and the expansion terms follow this same decreasing trend in magnitude. The largest contribution to Equation (2) arises from the expansion term $(\nabla \cdot \boldsymbol{u}_e)\epsilon$, and the smallest contribution arises from the divergence of the heat flux $\nabla \cdot \boldsymbol{Q}_e$.

The magnitude of the divergence of the heat flux is $\sim 3 \times 10^{-14}$ W m$^{-3}$ at 0.15 au and decreases to $\sim 7 \times 10^{-16}$ W m$^{-3}$ at 0.48 au. The divergence of the heat flux is positive at $r < 0.33$ au and negative at $r > 0.33$ au. It contributes $\sim 10\%$ to $\Xi$ at $r < 0.33$ au. The radial gradient of $\Xi$ steepens near 0.33 au, where the divergence of heat flux changes sign.

The pressure strain term and the advection term are approximately equal in magnitude at $r > 0.35$ au yet with opposite signs. The advection term and the pressure strain term thus largely cancel each other so that the thermal balance is mainly determined by the competition between the remaining terms: the divergence of the heat flux and the expansion term.

$\Xi$ is positive across all explored heliocentric distances within the error bars. It drops below the contribution from the expansion term at $r > 0.3$ au.

Figure 3 shows our analysis of the polytropic index. We display the logarithm of the total scalar electron pressure as a function of the logarithm of the total electron density. We fit a straight line of the form shown in Equation (21). From the best-fitted line, we obtain $\gamma_e = 1.18$ and $C = -18.8$.

## 5. Discussion

To the best of our knowledge, this is the first work to present a description of the electron energetics in the solar wind based on the complete and exact second moment of the Boltzmann equation and application of the electron thermal energy budget to the inner heliosphere. The critical parameter that is evaluated in this paper is the irreversible electron thermal power density $\Xi$, which measures the deposition of the thermal energy density. If $\Xi = 0$, then the electron thermal budget is fully governed by free-streaming effects such as the advection of the thermal energy density by the bulk flow, divergence of the heat flux, pressure strain, and the expansion of the solar wind without additional deposition or removal of thermal energy density. However, Figure 2 shows that $\Xi$ is on average positive from 0.15 to 0.47 au. Therefore, the free streaming alone is not sufficient to maintain the thermal energy budget of the electrons in the solar wind.

This finding suggests that additional processes provide thermal energy on average across all measured radial distances to balance the thermal budget. The magnitude of $\Xi$ decreases with radial distance, suggesting that less heating is required to maintain the energy balance with greater heliocentric distance.

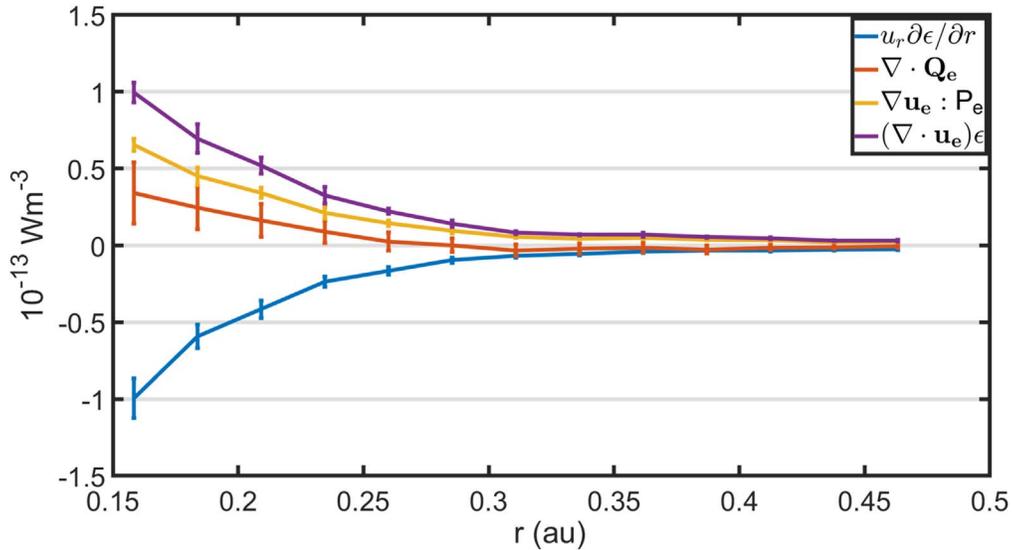

**Figure 1.** Thermal energy budget of solar wind electrons as a function of radial distance. The blue line represents the advection of the thermal energy by the bulk flow. The red line represents the divergence of the heat flux. The yellow line represents the pressure strain term. The purple line represents the expansion term.

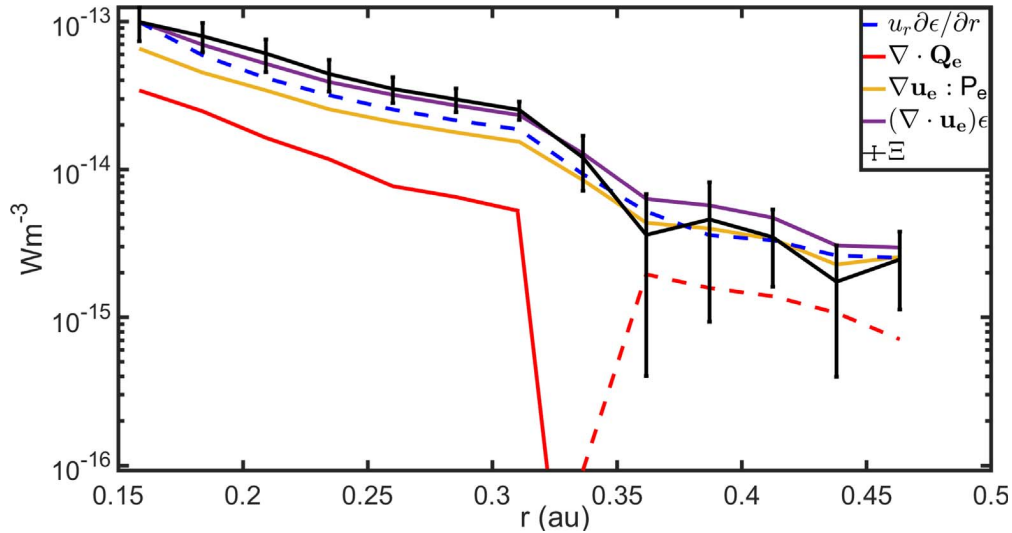

**Figure 2.** Thermal energy budget of solar wind electrons as a function of radial distance. The color scheme is the same as in Figure 1. The black line represents $\Xi$ with cumulative error bars. We use solid lines when a quantity is positive and dashed lines when a quantity is negative.

The divergence of the heat flux is positive below 0.33 au and negative beyond 0.33 au. Our findings of the degradation of the heat flux at radial distances greater than 0.33 au are in agreement with findings from Helios (Štverák et al. 2015). The change in nature of the heat flux suggests the action of a mechanism that shapes the VDF so that the heat flux decreases more slowly than $\propto r^{-2}$ at distances below 0.33 au and faster than $\propto r^{-2}$ at distances beyond. PSP observations have shown that, at closest distances less than 0.2 au from the Sun, there is only a slight or even no halo signature. However, the halo signature becomes prominent with radial distances greater than 0.3 au (Halekas et al. 2021; Abraham et al. 2022), which is potentially related to the observed heat-flux behavior.

Equation (2) does not encapsulate the full energy conservation of the solar wind because it does not account for the electromagnetic energy, the electron bulk kinetic energy, and the energy associated with other particle species. Equation (2) is accurate and complete in its description of the thermal energy budget of the electron species, though.

We identify three potential mechanisms that could possibly provide external thermal energy sources or sinks to the electrons in the form of nonzero $\Xi$: turbulent heating, instabilities, and collisions. The turbulent cascade transfers energy from large scales to kinetic scales, where kinetic processes dissipate the energy in the form of heat (Tu & Marsch 1995; Breech et al. 2009; Schekochihin et al. 2009; Bruno & Carbone 2013; Goldstein et al. 2015; Livadiotis 2019; Franci et al. 2022). This form of turbulent dissipation leads to an irreversible deposition of thermal energy. Empirical studies suggest that a significant amount (∼40%) of the turbulent energy is dissipated into electrons in the inner heliosphere (Cranmer et al. 2009; Shoda et al. 2021). Such a contribution, if dominant in the overall irreversible power density, leads to a positive value of $\Xi$ as observed. Recent observations of the turbulence spectrum have shown a substantial change in the spectral index of the inertial-range turbulence power spectral density with heliocentric distance from $-3/2$ for $r \lesssim 0.3$ au to $-5/3$ for $r \gtrsim 0.3$ au (Chen et al. 2020). This change potentially influences the deposition of turbulent energy into the electrons and may thus explain the observed change in $\Xi$ around 0.3 au. In this case, turbulent electron heating can be considered a potentially significant contributor to the electron thermal energy budget at the heliocentric distances explored in our study. The typically considered dissipation channels for turbulent heating arise from kinetic Alfvén waves (Bale et al. 2005; Schekochihin et al. 2009; Malara et al. 2019; Zhou et al. 2022), magnetic switchbacks (Bale et al. 2021; Shoda et al. 2021), shear that arises from stream–stream interaction (Coleman 1968), or nonlinear dissipation in current sheets (Sundkvist et al. 2007; Agudelo Rueda et al. 2021; Bandyopadhyay et al. 2021). However, it is outside the scope of this study to distinguish the kinetic channels of dissipation for turbulent heating.

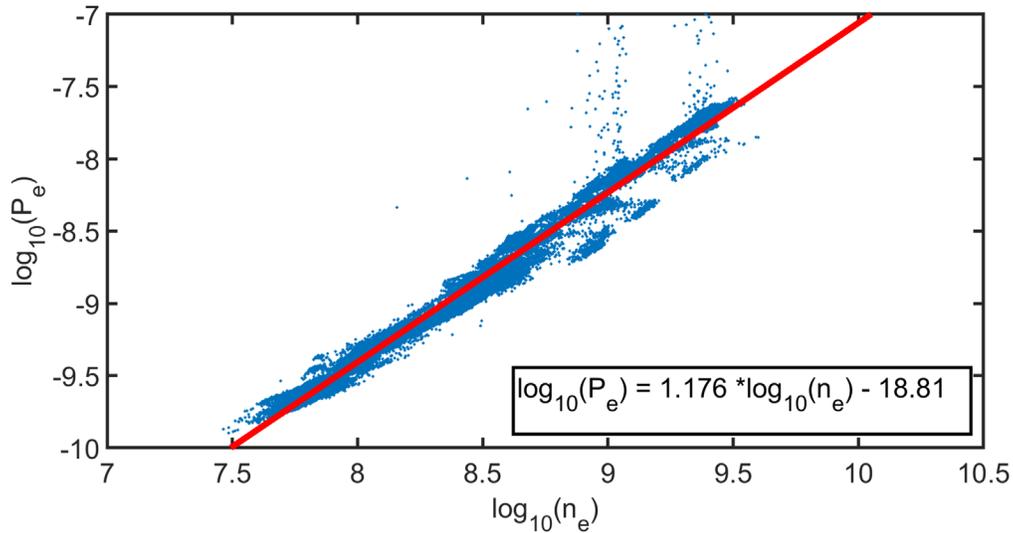

**Figure 3.** Determination of the polytropic relationship of solar wind electrons. Blue dots show the total thermal electron pressure as a function of the total electron density in double-logarithmic space. The red line represents the best-fitted straight line given in the box in the bottom right-hand corner.

Deviations from thermodynamic equilibrium can create and drive fluctuations in the electric and magnetic fields in the form of plasma instabilities. As these instabilities grow, they interact with the electrons, leading to a change in the electron VDF such that the drivers of the instability are reduced and the VDF achieves a stable state (Feldman et al. 1976; Schwartz 1980; Štverák et al. 2015; Verscharen et al. 2019; Jeong et al. 2020, 2022a; Verscharen et al. 2022). Instabilities driven by temperature anisotropy act on the electron VDF and drive the VDF toward anisotropy. Likewise, instabilities driven by heat flux reduce the heat flux in the VDF (Gary et al. 1975; López et al. 2020). Instabilities generally transfer energy from the particles into the growing electromagnetic fluctuations. Therefore, we expect a reduction in thermal energy during the time of the growth of the instabilities. In our energetics framework, this would correspond to a negative contribution to $\Xi$. As we do not observe $\Xi < 0$ on average, our results suggest that electron-driven instabilities are not a dominant contributor to the average electron thermal energy budget through the irreversible electron thermal power density. We note, however, that this finding does not rule out the existence and relevance of electron-driven instabilities altogether. It is merely a statement regarding their dominance in the energy budget according to Equation (2). Since the magnitude of $\Xi$ decreases with radial distance, instabilities may be a key driver of the thermal energy budget at greater heliocentric distances (see also Jeong et al. 2022b).

Collisions can heat or cool the electrons, leading to a nonzero $\Xi$ term. For example, collisions between hot protons and cold electrons heat the electrons and vice versa (Salem et al. 2003). However, the consistent difference in temperature between the protons and the electrons in the solar wind shows that they are not in thermal equilibrium, which indicates that the collisional coupling between electrons and protons is weak (Feldman et al. 1975). Recent work on the heat flux in the inner heliosphere shows that collisions are not the dominant heat-flux regulation mechanism near the Sun (Bale et al. 2013; Halekas et al. 2021). Indeed, kinetic simulations and observations indicate that collisions contribute to the shaping of the VDF at small heliocentric distances $\lesssim 10 R_s$, but they are unlikely to play a major role in the balancing of the thermal energy density (Landi et al. 2012; Štverák et al. 2015; Jeong et al. 2022a).

For our data set, we find an effective polytropic index of $\gamma_e = 1.18$. This effective polytropic index is the result of the combined action of all terms accounted for in Equation (2). Its observed value is slightly greater than for an isothermal plasma ($\gamma_e = 1$). However, the measured value is below the polytropic index prediction of 1.23–1.29 in the isopoly model of the solar wind (Dakeyo et al. 2022). Nonetheless, our observed value and the isopoly model prediction are in general agreement in that the value lies between the isothermal and adiabatic condition. Our measured effective polytropic index agrees well with the predicted polytropic index of 1.17 from the Scudder & Olbert (1979) model.

## 6. Conclusions

We present a framework in which we apply fit results from spacecraft measurements to evaluate the complete and exact electron thermal energy budget based on the Boltzmann equation. We evaluate the average radial evolution of the thermal energy density budget in the inner heliosphere by separating the thermal energy balance between reversible and irreversible processes.

Under the assumption of a steady state, free-streaming effects of the electrons alone cannot account for the overall electron thermal energy budget across all measured distances. We find that the irreversible thermal power density $\Xi$ remains positive from 0.15 to 0.47 au. This result suggests the presence of an energy source that supplies the electrons with thermal energy across all radial distances explored in this study. We identify three mechanisms that are potentially responsible for a nonzero $\Xi$: turbulence, instabilities, and collisions. Turbulent heating is most likely the main driver responsible for the $\Xi > 0$ on average.

The pressure strain term is of the same magnitude as the advection term in our energy balance. The most dominant term, however, is the expansion term up to a distance of 0.33 au. The divergence of the heat flux only makes ∼10% contribution to the energy balance at distances between 0.15 and 0.33 au. The divergence of the heat flux is positive in this distance range. Beyond 0.33 au, however, the divergence of the heat flux is negative, which agrees with previous observations from Helios beyond 0.3 au (Štverák et al. 2015). This finding indicates that there is fundamental change in processes that define the shape of the electron VDF at around 0.33 au. We also find that the effective electron polytropic index is 1.18 as a result of all contributions to Equation (2).

After quantifying the thermal energy budget of the electrons, the next natural question is the examination of the irreversible kinetic processes that lead to the observed deposition of electron thermal energy. In the future, it would be worthwhile to identify and quantify the processes that provide the thermal energy across different distances. Likewise, it would be interesting to identify and quantify the processes that define the sign of the divergence of heat flux within and beyond 0.33 au. A complex interplay of heat-flux generating processes (e.g., double-adiabatic focusing) and heat-flux destroying processes (e.g., instabilities) is likely responsible for this behavior.

In the future, alignments between PSP, Solar Orbiter, and potentially other spacecraft will help to measure solar wind plasma from the same source at different distances. This analysis will help us quantify the temporal variability without having to rely on the statistical consistency of multiple wind streams.

We acknowledge the NASA Parker Solar Probe Mission and the SWEAP team led by Justin Kasper for the use of data. The authors are grateful to the PSP instrument teams for producing and making the data used in this study publicly available. The SWEAP level-3 data are obtained from http://sweap.cfa.harvard.edu/pub/data/sci/sweap/spe/L3/spe_sf0_pad/. J.B.A. is supported by the Science and Technology Facilities Council (STFC) grant ST/T506485/1. D.V. is supported by the STFC Ernest Rutherford Fellowship ST/P003826/1. C.J.O and D.V. receive support from the STFC Consolidated Grants ST/S000240/1 and ST/W001004/1. R.T.W. is supported by STFC Consolidated Grant ST/V006320/1. J.A.A.R. is supported by NASA grant 80NSSC21K2048 and NSF grant 2142430. S.-Y.J. is supported by STFC grant ST/W000369/1. This work was discussed at the "Joint Electron Project (JEP)" at MSSL and at the ISSI Team "Heliospheric Energy Budget: From Kinetic Scales to Global Solar Wind Dynamics" led by M. E. Innocenti and A. Tenerani.

## ORCID iDs

Joel B. Abraham https://orcid.org/0000-0002-6305-3252
Daniel Verscharen https://orcid.org/0000-0002-0497-1096
Robert T. Wicks https://orcid.org/0000-0002-0622-5302
Jeffersson A. Agudelo Rueda https://orcid.org/0000-0001-5045-0323
Christopher J. Owen https://orcid.org/0000-0002-5982-4667
Georgios Nicolaou https://orcid.org/0000-0003-3623-4928
Seong-Yeop Jeong https://orcid.org/0000-0001-8529-3217

## References

Abraham, J. B., Owen, C. J., Verscharen, D., et al. 2022, ApJ, 931, 118
Agudelo Rueda, J. A. 2022, PhD thesis, UCL (Univ. College London)
Agudelo Rueda, J. A., Verscharen, D., Wicks, R. T., et al. 2021, JPlPh, 87, 905870228
Agudelo Rueda, J. A., Verscharen, D., Wicks, R. T., et al. 2022, ApJ, 938, 4
Bale, S. D., Horbury, T. S., Velli, M., et al. 2021, ApJ, 923, 174
Bale, S. D., Kellogg, P. J., Mozer, F. S., Horbury, T. S., & Reme, H. 2005, PhRvL, 94, 215002
Bale, S. D., Pulupa, M., Salem, C., Chen, C. H. K., & Quataert, E. 2013, ApJL, 769, L22
Bandyopadhyay, R., Chasapis, A., Matthaeus, W. H., et al. 2021, PhPl, 28, 112305
Breech, B., Matthaeus, W. H., Cranmer, S. R., Kasper, J. C., & Oughton, S. 2009, JGRA, 114, A09103
Bruno, R., & Carbone, V. 2013, LRSP, 10, 2
Case, A. W., Kasper, J. C., Stevens, M. L., et al. 2020, ApJS, 246, 43
Chandrasekhar, S. 1967, An Introduction to the Study of Stellar Structure (New York: Dover Publications)
Chen, C. H. K., Bale, S. D., Bonnell, J. W., et al. 2020, ApJS, 246, 53
Coleman, P. J. J. 1968, ApJ, 153, 371
Cranmer, S. R., Matthaeus, W. H., Breech, B. A., & Kasper, J. C. 2009, ApJ, 702, 1604
Dakeyo, J. B., Maksimovic, M., Démoulin, P., Halekas, J., & Stevens, M. L. 2022, arXiv:2207.03898
Feldman, W., Asbridge, J., Bame, S., Montgomery, M., & Gary, S. 1975, JGR, 80, 4181
Feldman, W. C., Asbridge, J. R., Bame, S. J., Gary, S. P., & Montgomery, M. D. 1976, JGR, 81, 2377
Feldman, W. C., Asbridge, J. R., Bame, S. J., & Gosling, J. T. 1979, JGR, 84, 7371
Franci, L., Papini, E., Micera, A., et al. 2022, ApJ, 936, 27
Gary, S. P., Feldman, W. C., Forslund, D. W., & Montgomery, M. D. 1975, JGR, 80, 4197
Goldstein, M. L., Wicks, R. T., Perri, S., & Sahraoui, F. 2015, RSPTA, 373, 20140147
Halekas, J. S., Berčič, L., Whittlesey, P., et al. 2021, ApJ, 916, 16
Halekas, J. S., Whittlesey, P. L., Larson, D. E., et al. 2021, A&A, 650, A15
Jeong, S.-Y., Abraham, J., Verscharen, D., et al. 2022b, ApJL, 926, L26
Jeong, S.-Y., Verscharen, D., Vocks, C., et al. 2022a, ApJ, 927, 162
Jeong, S.-Y., Verscharen, D., Wicks, R. T., & Fazakerley, A. N. 2020, ApJ, 902, 128
Kasper, J. C., Abiad, R., Austin, G., et al. 2016, SSRv, 204, 131
Kuhn, S., Kamran, M., Jelić, N., et al. 2010, in AIP Conf. Ser. 1306, New Frontiers in Advanced Plasma Physics, ed. B. Eliasson & P. K. Shukla (Melville, NY: AIP), 216
Landi, S., Matteini, L., & Pantellini, F. 2012, ApJ, 760, 143
Landi, S., Matteini, L., & Pantellini, F. 2014, ApJ, 790, L12
Lemaire, J., & Scherer, M. 1971, JGR, 76, 7479
Lie-Svendsen, Ø., Hansteen, V. H., & Leer, E. 1997, JGR, 102, 4701
Livadiotis, G. 2019, ApJ, 887, 117
López, R. A., Lazar, M., Shaaban, S. M., Poedts, S., & Moya, P. S. 2020, ApJ, 900, L25
Maksimovic, M., Gary, S. P., & Skoug, R. M. 2000, JGR, 105, 18337
Maksimovic, M., Zouganelis, I., Chaufray, J.-Y., et al. 2005, JGRA, 110, A09104
Malara, F., Nigro, G., Valentini, F., & Sorriso-Valvo, L. 2019, ApJ, 871, 66
Marsch, E. 2006, LRSP, 3, 1
Nicolaou, G., Livadiotis, G., Wicks, R. T., Verscharen, D., & Maruca, B. A. 2020, ApJ, 901, 26
Ogilvie, K. W., & Scudder, J. D. 1978, JGR, 83, 3776
Parker, E. 2010, in AIP Conf. Proc. 1216, Twelfth Int. Solar Wind Conf. (Melville, NY: AIP), 3
Parker, E. N. 1963, Interplanetary Dynamical Processes (New York: Interscience)
Pilipp, W. G., Miggenrieder, H., Mühläuser, K.-H., Rosenbauer, H., & Schwenn, R. 1990, JGR, 95, 6305
Salem, C., Hubert, D., Lacombe, C., et al. 2003, ApJ, 585, 1147
Schekochihin, A. A., Cowley, S. C., Dorland, W., et al. 2009, ApJS, 182, 310

Schwartz, S. J. 1980, RvGSP, 18, 313
Scudder, J. D., & Olbert, S. 1979, JGR, 84, 6603
Shoda, M., Chandran, B. D. G., & Cranmer, S. R. 2021, ApJ, 915, 52
Smith, H. M., Marsch, E., & Helander, P. 2012, ApJ, 753, 31
Štverák, Š., Maksimovic, M., Trávníček, P. M., et al. 2009, JGRA, 114, A05104
Štverák, Š., Trávníček, P., Maksimovic, M., et al. 2008, JGRA, 113, A03103
Štverák, Š., Trávníček, P. M., & Hellinger, P. 2015, JGRA, 120, 8177
Sundkvist, D., Retinò, A., Vaivads, A., & Bale, S. D. 2007, PhRvL, 99, 025004
Totten, T. L., Freeman, J. W., & Arya, S. 1995, JGR, 100, 13
Tu, C.-Y., & Marsch, E. 1995, SSRv, 73, 1
Verscharen, D., Chandran, B. D. G., Boella, E., et al. 2022, FrASS, 9, 951628
Verscharen, D., Chandran, B. D. G., Jeong, S.-Y., et al. 2019, ApJ, 886, 136
Vocks, C., & Mann, G. 2003, ApJ, 593, 1134
Whittlesey, P. L., Larson, D. E., Kasper, J. C., et al. 2020, ApJS, 246, 74
Yang, Y., Matthaeus, W. H., Roy, S., et al. 2022, ApJ, 929, 142
Zhou, M., Liu, Z., & Loureiro, N. F. 2022, arXiv:2208.02441